\documentclass[aps,prd, preprintnumbers ,floats,twocolumn,superscriptaddress,nofootinbib]{revtex4}
\usepackage{graphicx}
\usepackage{dcolumn}
\usepackage{bm}
\usepackage[english]{babel}
\usepackage[latin1]{inputenc} 
\usepackage{graphicx}
\usepackage{epstopdf}
\usepackage{amsfonts}
\usepackage{color}
\usepackage{epsfig}
\usepackage{mathrsfs}
\usepackage{verbatim} 
\usepackage{amsmath,amssymb}
\usepackage{subfigure}
\usepackage{hyperref}
\usepackage{framed}
\usepackage{lipsum}
\usepackage[svgnames]{xcolor}
\definecolor{shadecolor}{named}{LightGrey}

\newcommand{\lto}[1]{\longrightarrow#1}

\renewcommand{\(}{\left(}
\renewcommand{\)}{\right)}
\renewcommand{\[}{\left[}
\renewcommand{\]}{\right]}

\newcommand{\superscript}[1]{\ensuremath{^{\textrm{#1}}}}

\begin{document}

\graphicspath{{figure/}}
\selectlanguage{english}

\title{The Anatomy of a Scientific Rumor}

\author{M. De Domenico}
 \affiliation{School of Computer Science, University of Birmingham, United Kingdom}

\author{A. Lima}
 \affiliation{School of Computer Science, University of Birmingham, United Kingdom}

\author{P. Mougel}
 \affiliation{School of Computer Science, University of Birmingham, United Kingdom}
 \affiliation{Universit\'e de Lyon, INSA-Lyon, Villeurbanne, France}

\author{M. Musolesi}
 \affiliation{School of Computer Science, University of Birmingham, United Kingdom}


\begin{abstract}
The announcement of the discovery of a Higgs boson-like particle at CERN will be remembered as one of the milestones of the scientific endeavor of the 21\superscript{st} century.
In this paper we present a study of information spreading processes on Twitter before, during and after the announcement of the discovery of a new particle with the features of the elusive Higgs boson on 4\superscript{th} July 2012. 
We report evidence for non-trivial spatio-temporal patterns in user activities at individual and global level, such as tweeting, re-tweeting and replying to existing tweets.
We provide a possible explanation for the observed time-varying dynamics of user activities during the spreading of this scientific ``rumor''. We model the information spreading in the corresponding network of individuals who posted a tweet related to the Higgs boson discovery. Finally, we show that we are able to reproduce the global behavior of about 500,000 individuals with remarkable accuracy.
\end{abstract}

\maketitle

\flushbottom


The Higgs boson, whose existence has been hypothesized in 1964\,\cite{higgs1964broken}, has gained the title of the most elusive particle in modern science. The search for its existence has been among the top research priorities of the particle physics community for nearly 50 years. 2012 will be probably remembered as one of the most important years in this century for physics: on 4\superscript{th} July 2012 the ATLAS and CMS collaborations, two international experiments involved in the search for the Higgs boson, announced the results of the discovery of a new particle with the features of the elusive Higgs boson, the missing component of the Standard Model.

The elusive nature of the Higgs boson required the development of a new generation of large-scale experimental facilities, resulting in the construction of the Large Hadron Collider (LHC) at CERN, in Gen\'eve (Switzerland), the largest and most powerful particle accelerator ever built. The other detector able to find hints about the existence of the Higgs boson is the Tevatron at Batavia, IL (USA). 
The association of the Higgs boson to the idea of the final understanding of our Universe and the possibility of the Grand Unified Theory\,\cite{cabibbo1979bounds,langacker1981grand,ellis1989higgs,amaldi1991comparison} is likely to be responsible for the huge popularity of this research project in both academic and non-academic circles. Indeed, the interest from both specialized and popular media increased after the ``God particle'' nickname was assigned to the Higgs boson \cite{lederman1993god}.

The announcement of this discovery was the first of this kind in the era of global online social media, such as Twitter: the entire world followed and discussed the news and updates through them, commenting and providing personal views about the event. All this information is publicly generated online and represents an extremely interesting source of data for analyzing the global dynamics of this scientific rumor around the world.

On 2\superscript{nd} July, initial results were presented by the Tevatron team, but they were not sufficient to claim a scientific discovery. The statistical significance of all the combined analyses was 2.9 sigma, equivalent to a 1-in-550 chance that the signal was due to a statistical fluctuation~\cite{tevatron2012updated}. Although of remarkable importance for the scientific community, such an announcement had a weak impact on the general public. 
Following this, there was a strong expectation, accompanied by rumors, for the corresponding results from the CERN teams. An unofficial video was even leaked during those days~\cite{leakedvideo2012}.
The spreading of these rumors about a possible discovery attracted the interest of media, also outside the academic community, until the official day of the announcement on  4\superscript{th} July during the International Conference on High-Energy Physics 2012 in Melbourne, Australia.

We can summarize the events before and after the discovery of the boson, dividing them into 4 different periods:
\begin{itemize}
\item \textbf{Period I:} Before the announcement on 2\superscript{nd} July, there were some rumors about the discovery of a Higgs-like boson at Tevatron;
\item \textbf{Period II:} On 2\superscript{nd} July at 1 PM GMT, scientists from CDF and D0 experiments, based at Tevatron, presented results indicating that the Higgs particle should have a mass between 115 and 135~GeV/c$^{2}$ (corresponding to about 123-144 times the mass of the proton)~\cite{tevatron2012updated};
\item \textbf{Period III:} After 2\superscript{nd} July and before 4\superscript{th} of July there were many rumors about the Higgs boson discovery at LHC~\cite{leakedvideo2012};
\item \textbf{Period IV:} The main event was the announcement on 4\superscript{th} July at 8 AM GMT by the scientists from the ATLAS and CMS experiments, based at CERN, presenting results indicating the existence of a new particle, compatible with the Higgs boson, with mass around 125~GeV/c$^{2}$~\cite{atlas2012updated,cms2012updated}. After 4\superscript{th} July, popular media covered the event.
\end{itemize}

In this paper, we present the anatomy of the spreading of this scientific rumor by following and analyzing the related Twitter user activity during and after the announcement. More specifically, we consider the messages posted in Twitter about this discovery between 1\superscript{st} and 7\superscript{th} July 2012. 
We report evidence for non-trivial spatio-temporal patterns in user activities at individual and global level, as tweeting, re-tweeting or replying to existing tweets. Well-defined trends can be associated to different periods of time. Abrupt changes can be linked to the key events around the announcement.
We analyze the activity patterns of the individuals that tweeted about this discovery over the period taken into consideration. We propose a model for the information spreading over the Twitter network, assuming memoryless individuals where the activation process is driven by social reinforcement at neighborhood level. Finally, we show that we are able to reproduce the global behavior of more than 500,000 individuals with remarkable accuracy.

\section*{Results}

\subsection*{Overview of the Dataset}

Our dataset consists of messages posted on the Twitter social network, crawled by means of the Application Programming Interface (API) made available by the service itself.  We collected tweets sent between 00:00 AM, 1\superscript{st} July 2012 and
11:59 PM, 7\superscript{th} July 2012 containing at least one of the
following keywords or hashtags: \texttt{lhc}, \texttt{cern}, \texttt{boson},
\texttt{higgs} (see \emph{Methods}). The final amount of tweets we analyzed was $985,590$.  
Hence, we built the corresponding social network of the authors of the tweets: 
the resulting graph is composed of 456,631 nodes and 14,855,875 directed edges.
Nodes correspond to the authors of the tweets and edges represent the followee/follower relationships between them.
 We discarded 70,838 users from the original dataset 
containing 527,469 users because of the non accessibility of the list of their followees and followers due to privacy settings. 
Twitter users can specify their location by filling the \textit{Location} field
of their profile, on optional basis and at different levels of granularity (e.g.,
United States, New York, Chelsea, etc.). We use this information, when available, to
assign a geographic position to each tweet: the resulting number of geo-located tweets is 632,027 (see \emph{Methods}).

\begin{figure}[!t]
\includegraphics[width=9cm]{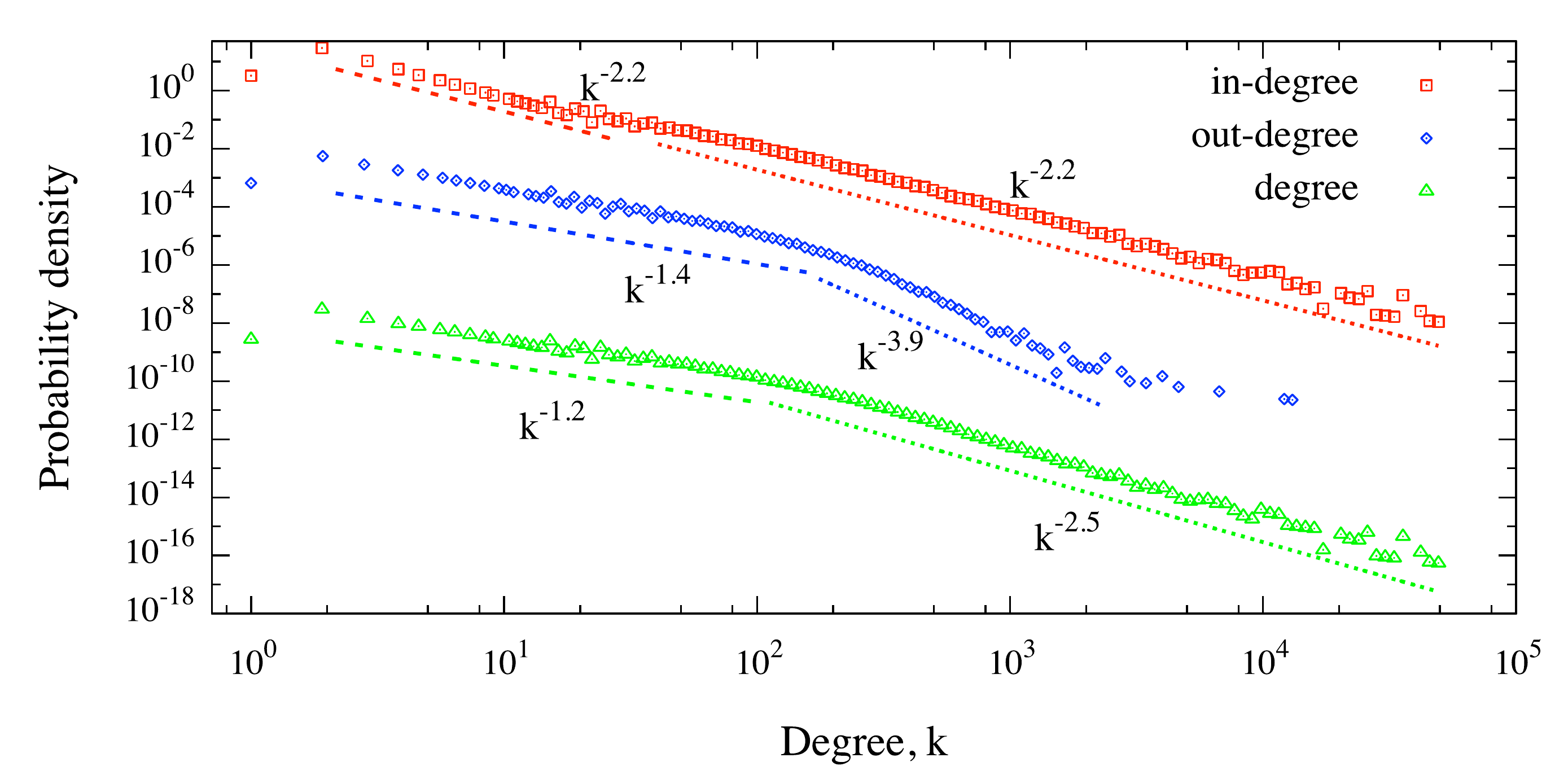}
\caption{Probability density of in-degree, out-degree and total degree of the nodes that tweeted about the Higgs boson. The corresponding distributions have been shifted along the $y$ axis to put in evidence their structure. Dashed lines are shown for guidance only.}
\label{fig:degreedistr}
\end{figure}

In Fig.\,\ref{fig:degreedistr} we show the distributions of the in-degree, out-degree and total degree of the users that tweeted about the Higgs boson.
Intriguingly, the underlying topology is not trivial. The out-degree distribution shows a power-law scaling with two different regimes: $P(k_{\text{out}})\propto k_{\text{out}}^{-1.4}$ and $P(k_{\text{out}})\propto k_{\text{out}}^{-3.9}$, with crossover for $k_{\text{out}}\approx200$, which indicates that very few users follow more than a few hundred users. Conversely, the in-degree distribution shows a different behavior: for in-degree smaller than $k_{\text{in}}\approx100$ the scaling relation is not satisfied, whereas above this threshold the network exhibits a power-law scaling $P(k_{\text{in}})\propto k_{\text{in}}^{-2.2}$.
A standard method to uncover the presence of correlations in the network is to investigate the assortative mixing of its nodes \cite{newman2002assortative,newman2003mixing}. In fact, the nodes in the network with a large number of links may tend to be connected to other nodes with many connections (assortative mixing with positive assortative index) or to other nodes with a few connections (disassortative mixing with negative assortative index). In both cases, the network shows degree correlations resulting in an assortative index different from zero, at variance with an uncorrelated network where this index is close to zero. In our case study we find a value of about -0.14, indicating the presence of correlations in the network, with disassortative mixing of users.

\subsection*{Spatio-temporal Analysis}
\label{sec-temporal}

In this section, we investigate both spatial and temporal features of the observed data, i.e., user activity on Twitter before, during and after the main event on 4\superscript{th} July 2012. More specifically, we focus our attention on the study of user behavior by considering two different analyses: the first one is performed at a global (macroscopic) level, while the second one is performed at an individual (microscopic) level.

\begin{figure}[!t]
\includegraphics[width=9cm]{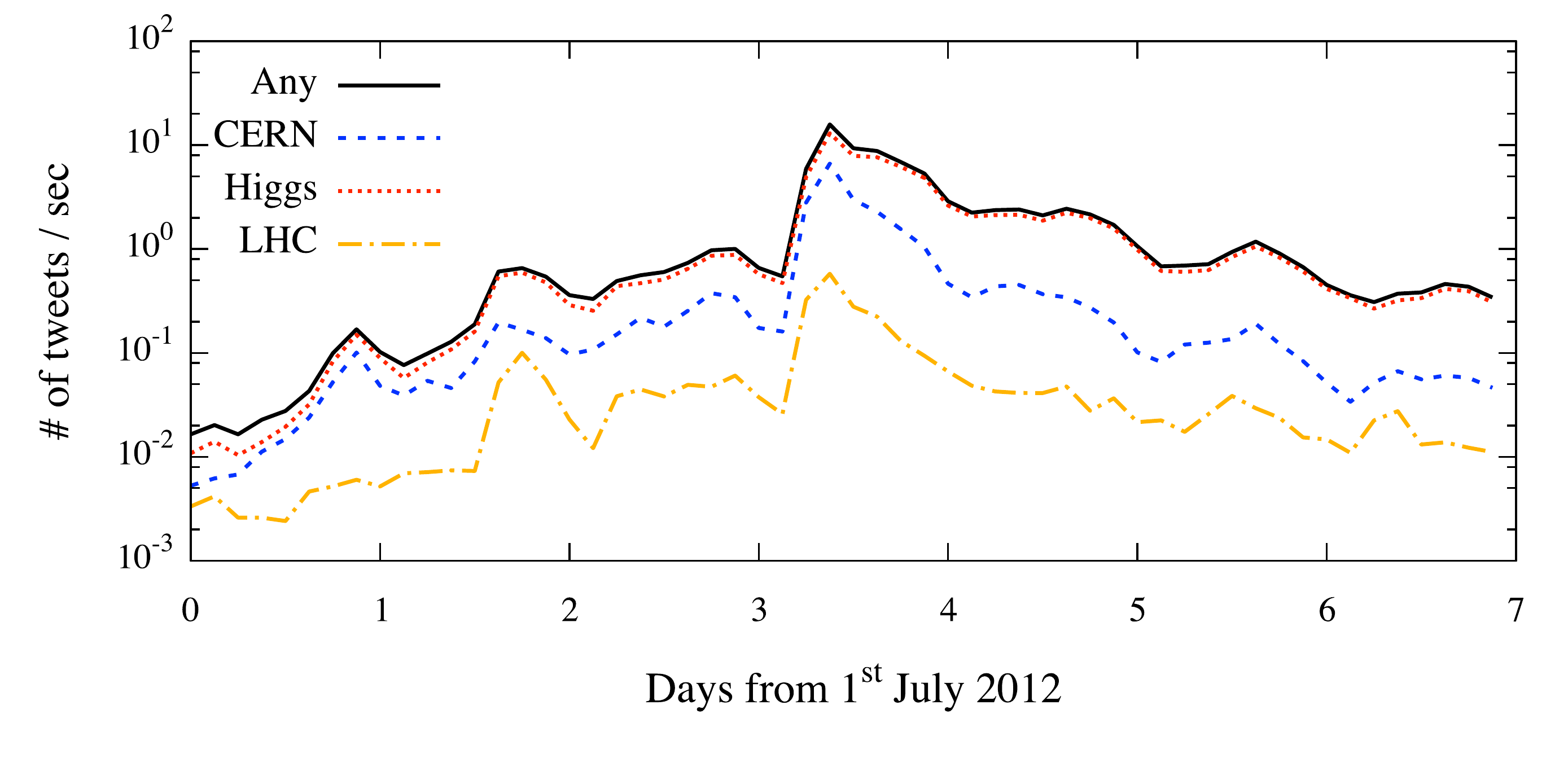}
\caption{Number of tweets per second as a function of time during the period of data collection. The curves correspond to tweets containing only the \texttt{CERN}, \texttt{Higgs}, \texttt{LHC} keywords and at least one of them, respectively.}
\label{fig:flow}
\end{figure}

\vspace{0.5truecm}\noindent\textbf{Macroscopic Level.} We consider the entire set of individuals as a large-scale complex system of interacting entities, and we analyze the dynamics at a macroscopic level of such a system by inspecting spatio-temporal patterns of consecutive tweets. The inter-tweets time (space) is defined by the temporal delay (spatial distance) between two consecutive tweets posted by \emph{any} user in the network. 

In Fig.\,\ref{fig:flow} we show the evolution of the rate of tweets containing the \texttt{CERN}, \texttt{Higgs}, and \texttt{LHC} keywords. The rate shows a rapidly increasing trend up to the day of the announcement of the CERN teams, after which it slowly decreases. It is worth noting that, when all the keywords are considered, the rate of tweets increases from approximately 36 tweets/hour at the beginning of Period I up to about 36,000 tweets/hour at the beginning of Period IV. The rumors anticipating the presentation of results at Tevatron caused the initial spreading of tweets about the Higgs boson. This was further sustained by the subsequent comments to these initial postings and the rumors about the results to be presented by the scientists belonging to the ATLAS and CMS experiments. 
During a few hours after the announcement of the discovery, the rate increased by more than one order of magnitude, while it slowly decreased in the following days.

\begin{figure*}[!t]
\includegraphics[width=17.5cm]{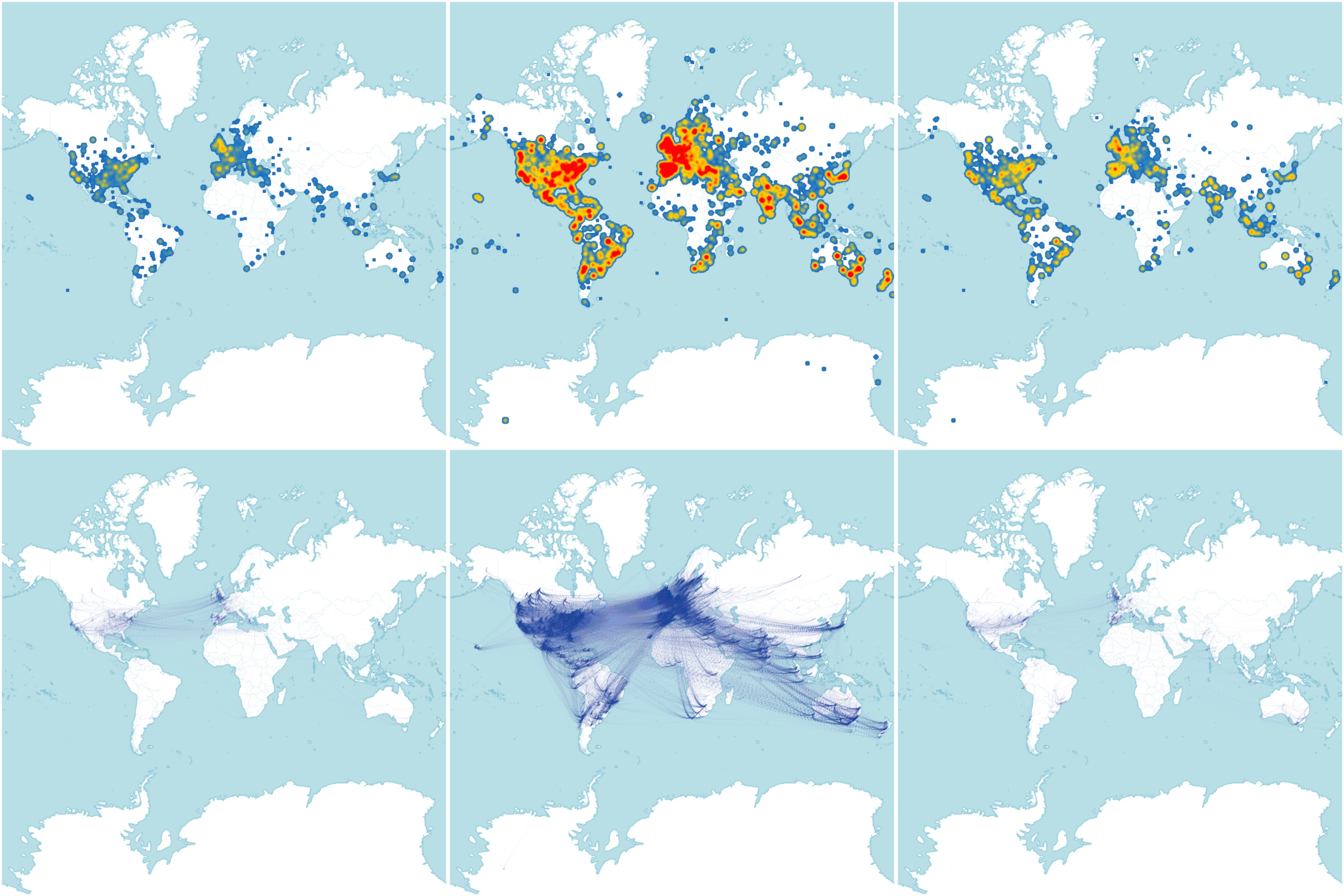}
\caption{Top: heatmap for the density of tweets before (left panel), during (middle panel) and after (right panel) the main event on 4\superscript{th} July 2012. Bottom: corresponding networks of re-tweets between users. During the announcement, the Twitter activity is truly global, whereas before and after the announcement, the most active countries were European and American, due to the large presence of scientists in these geographic areas. Map data: $\copyright$ OpenStreetMap contributors, available under the Open Database License.}
\label{fig:world}
\end{figure*}

In the top panels of Fig.\,\ref{fig:world} we show the density of tweets before (left panel), during (middle panel) and after (right panel) the main event, on 4\superscript{th} July 2012. In the bottom panels in the same figure we show the corresponding networks of users built from re-tweets.

The impact of the announcement on 4\superscript{th} was truly global. Instead, before and after this main event the countries with a significant number of tweets were European, probably due to the fact that CERN is in Switzerland and the largest number of scientists working there are from Europe. A large number of tweets were also observed from the United States, which hosts a very large community of scientists. 

\begin{figure*}[!t]
   \includegraphics[width=\textwidth]{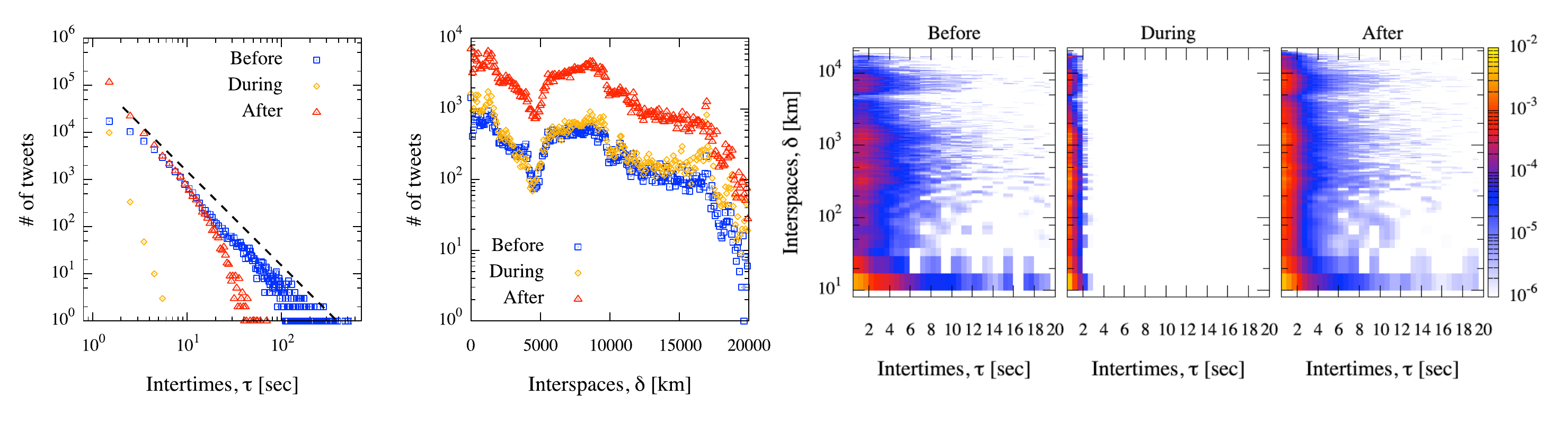}
 \caption{\textbf{First and second panels:} Global spatio-temporal activities of \emph{any} user in the social network. Number of entries for inter-tweets times (first panel) and inter-tweets spaces (second panel) between consecutive tweets, before, during and after the main event on 4\superscript{th} July. The dashed line indicates a power law $\sim\tau^{-2}$ and is for guidance only. \textbf{Third, fourth and fifth panels:} Joint probability density of inter-tweets times and inter-tweets spaces between consecutive tweets before (third panel), during (fourth panel) and after (fifth panel) the main event. In both cases, only the sub-set of geo-located tweets is considered.}
    \label{fig:interspacetimes}
\end{figure*} 

Our first goal is to gain insights into the spatial and temporal patterns of this complex geographic social network: in order to do so, we estimate the distance in time and space of tweets posted in the network. In Fig.\,\ref{fig:interspacetimes} we show the number of tweets as a function of the inter-tweets time (first panel) and the inter-tweets space (second panel) between two consecutive messages. In both cases, we show the distributions corresponding to the period before, during and after the main event on 4\superscript{th} July, respectively. While the distribution of inter-tweets spaces is the same regardless of the time window taken into consideration, the distribution of inter-tweets times in the three windows is very different. From a global point of view, this Twitter activity exhibits long tails before and after the main event, with a large number of tweets sent within a few seconds, and a small number sent within a few minutes. On the other hand, the dynamics of the process changes dramatically during the main event, when the inter-tweets time between consecutive tweets is likely to be less than two seconds and no more than six seconds, indicating a frenetic user activity. A deeper investigation at an individual level of this bursty behavior is presented in the next section.

In order to unveil the presence of spatio-temporal patterns of individuals with non-trivial relationships, in the last three panels of Fig.\,\ref{fig:interspacetimes} we show the joint probability density of inter-tweets times and inter-tweets spaces before (third panel), during (fourth panel) and after (fifth panel) the main event. Before the main event, consecutive tweets are mainly sent at local scales within less than half a minute interval, generally within 8 seconds: consecutive tweets are more likely to be sent by users living within 20\,km, although a significant number of tweets is still posted on larger inter-tweets spatial scales. The system dynamics changes dramatically during the main event: tweets from any part of the world are likely to be sent within 2 seconds without a specific spatial pattern. User activity now is frenetic and information is quickly spreading at any spatial scale. After the main event, the spatio-temporal dynamics tends to become similar to the activity before the main event, even if, in this case, users from any part of the world are still involved in the process, with no apparent prevalence of small or large inter-tweets spaces.

\vspace{0.5truecm}\noindent\textbf{Microscopic Level.} We now analyze the dynamics at a microscopic level (i.e., treating individuals separately) by inspecting inter-arrival times of activities as tweeting, replying and re-tweeting. In the following, the inter-activity time for user $u$ is defined by $\tau_{u}(i)=t_{u}(i+1)-t_{u}(i)$, where $t_{u}(i)$ and $t_{u}(i+1)$ indicate the times when user $u$ sent the $i-$th and the $i+1$-th tweets, respectively. 

\begin{figure*}[!t]
 \includegraphics[width=\textwidth]{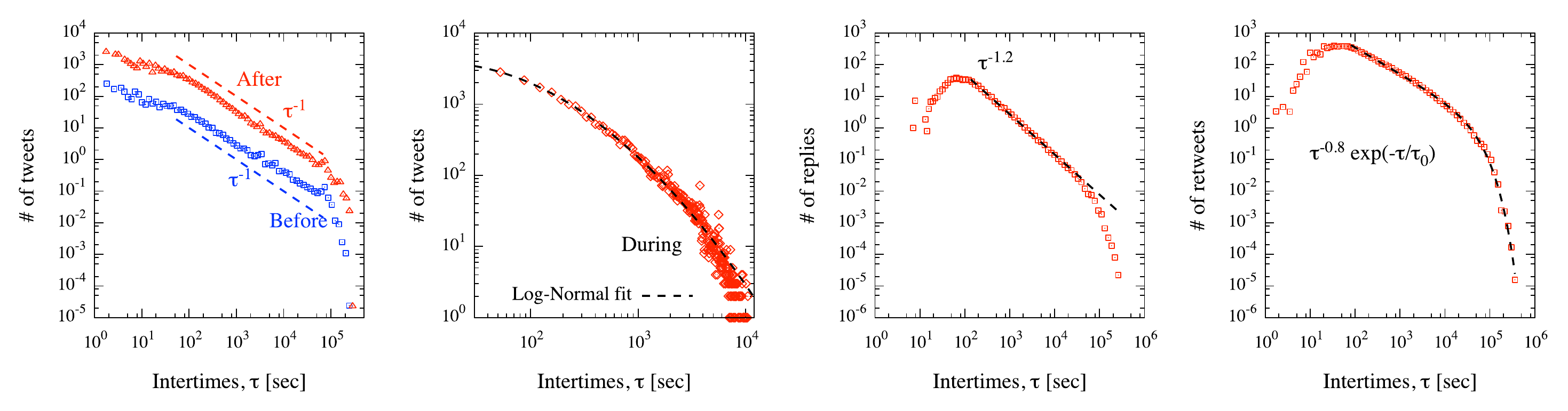}
 \caption{Activity inter-tweets times of users in the social network. \textbf{First panel:} Number of entries for inter-arrival times between consecutive tweets, before and after the main event on 4\superscript{th} July. Power scaling behavior is visible for certain ragnges of values, dashed lines are for guidance only. \textbf{Second panel:} Number of entries for inter-arrival times between consecutive tweets after the main event. The dashed curve corresponds to a lognormal fit. \textbf{Third panel:} Number of entries for inter-arrival times between replies. \textbf{Fourth panel:} Number of entries for inter-arrival times between re-tweets.}
   \label{fig:intertimes}
\end{figure*}

In the first two panels of Fig.\,\ref{fig:intertimes} we show the distribution of inter-tweets times $\tau$ (i.e., between consecutive tweets) during (first panel), before and after the main event (second panel). Intriguingly, before and after the main event, such distributions show power-law scaling of type $P(\tau)\propto \tau^{-\alpha}$, with $\alpha\approx 1$, over three decades of inter-tweets times, from the scale of one minute to the scale of one day.

Timing of user activities is usually modeled using a Poisson distribution. However, there is evidence that inter-tweets times between subsequent user actions follow a non-Poisson statistics, characterized by bursts of rapidly occurring events separated by long periods of inactivity~\cite{dewes2003analysis,masoliver2003continuous,karsai2012universal}. The bursty nature of user behavior has been recently attributed~\cite{barabasi2005origin} to decision-based queuing processes~\cite{cobham1954priority}, where individuals tend to act in response to some perceived priority. According to this model, the timing of tasks to be executed is heavy-tailed, with rapid responses in the majority of cases and a few responses with very long waiting times~\cite{barabasi2005origin}.  Moreover, it has been shown that bursty user activity patterns might have a remarkable impact on the spreading dynamics over complex networks: this dynamics might be related to the waiting time distribution but it is not sensitive to the network topology~\cite{min2011spreading}.
A similar dynamics is observed in our data: the distribution of inter-tweets times shown in the first panel of Fig.\,\ref{fig:intertimes} reflects the bursty nature of user activities on online social networks, where individuals are more likely to send several tweets in quick succession within a few minutes, followed by long periods of no or reduced activity, up to one day. 

Inter-tweets times distribution during the main event shows a very different behavior, more compatible with a log-normal law instead of a power-law scaling relationship. In this case, if $\tau$ is the random variable representing the time to the next tweet, the random variable $\log\tau$ is normally distributed with mean $\mu$ and standard deviation $\sigma$. If a user starts to tweet because he or she is triggered by tweets of users in his or her social neighborhood, the total number of tweets can pass a threshold value above which a cascading effect may occur in the network. If this is the case, we observe a random multiplicative~\cite{mitzenmacher2004brief} spreading of information, whose inter-tweets times are distributed following a log-normal law if the number of vertices involved in the cascade is large enough. The log-normal law with $\mu=5.627\pm0.008$ and $\sigma=1.742\pm0.006$ describes the observed activities during the main event with remarkable accuracy.

In the last two panels of Fig.\,\ref{fig:intertimes} we show the distributions of inter-arrival times for replies (third panel) and re-tweets (fourth panel) during the entire data collection period. Intriguingly, user activities are still characterized by bursty behavior. Inter-arrival times for replies follow a power-law $P(\tau)\propto \tau^{-1.2}$ from a few minutes up to one day: such a scaling can be explained with the existence of bursty behavior in the timing of user actions, previously discussed in the case of inter-tweets times before and after the main event. It is worth noting that the scaling exponent is larger for intervals between the original tweets and replies than for re-tweets. For temporal scales larger than one day, the power-law scaling is not present; an exponential cut-off does not model the observed decay.

The case of re-tweets deserves particular attention. From the time scale of a few minutes up to the time scale of a few hours, we observe the power-law scaling relationship $P(\tau)\propto \tau^{-0.8}$, with a cut-off on the time scale of one day. We model the data with a power-law with an exponential cut-off $P(\tau)\propto \tau^{-0.8}\exp(-\tau/\tau_{0})$, with cut-off scale $\tau_{0}\approx 11$\,hours. It is worth remarking that power-law scaling relationships with exponent $\alpha \leq1$ cannot be normalized and do not occur in nature unless the scaling deviates from power law after some threshold value, the cut-off scale, above which the distribution rapidly falls to zero. Even in such cases, phenomena exhibiting scaling exponents smaller than unity are very rare \cite{newman2005power}.

\subsection*{Rumor Spreading}

In this section, we investigate the dynamics of information spreading in the social network of users who twitted about the Higgs boson. 
Despite the fact that information spreading shares some general dynamical features with the spreading of diseases, their nature is deeply different. For instance, disease epidemics depends on the physical contacts between individuals and the different biological characteristics of both the infectious agent and the carrier, as well as many other factors \cite{anderson1992infectious}, whereas information can also be spread through non-physical contacts making use of communication infrastructures such as telephone, television and Internet \cite{myers2012information}. Information is very volatile and it is not subject to incubation periods: it is only worth spreading or not and this decision is made by individuals, unlike the case of disease spreading. In the last decade, the study of contagion dynamics, involving either information or disease transmission, has greatly benefited from key results in complex networks modeling \cite{pastor2001epidemicdyn,pastor2001epidemic, newman2002spread,newman2002email,moreno2002epidemic,keeling2005networks,leskovec2007cost,romero2011differences}: in fact, the structure of social relationships plays a fundamental role for any type of spreading dynamics \cite{newman2003structure,boccaletti2006complex,dorogovtsev2008critical}. If the underlying topology of the network is homogenous, the dynamics can be studied by adopting a mean-field approximation and the spreading occurs only if the rate of transmission of information exceeds an epidemic threshold. Conversely, heterogeneous structures like scale-free networks require heterogeneous mean-field approximation \cite{pastor2001epidemic,pastor2001epidemicdyn}, involving the single-site equation governing the time evolution of the relative density of ``infected'' vertices with given connectivity $k$, i.e., the probability that a vertex with degree $k$ is infected. Moreover, such networks have the peculiar property of facilitating the spreading of infections: in fact, if the corresponding degree distribution shows diverging second moment, then the epidemic threshold is zero independently from the degree correlations \cite{boguna2003absence}. Although mean-field approximations are fundamental tools to capture the main features of the spreading dynamics, particularly in the early stage, the models are less efficient when the finite size of the population becomes a significant factor. More recent approaches focus on the probability of transmission of individual vertices \cite{gomez2010discrete} and non-perturbative formulation of the heterogeneous mean-field approach \cite{gomez2011nonperturbative}.

\begin{figure}[!t]
\includegraphics[width=9cm]{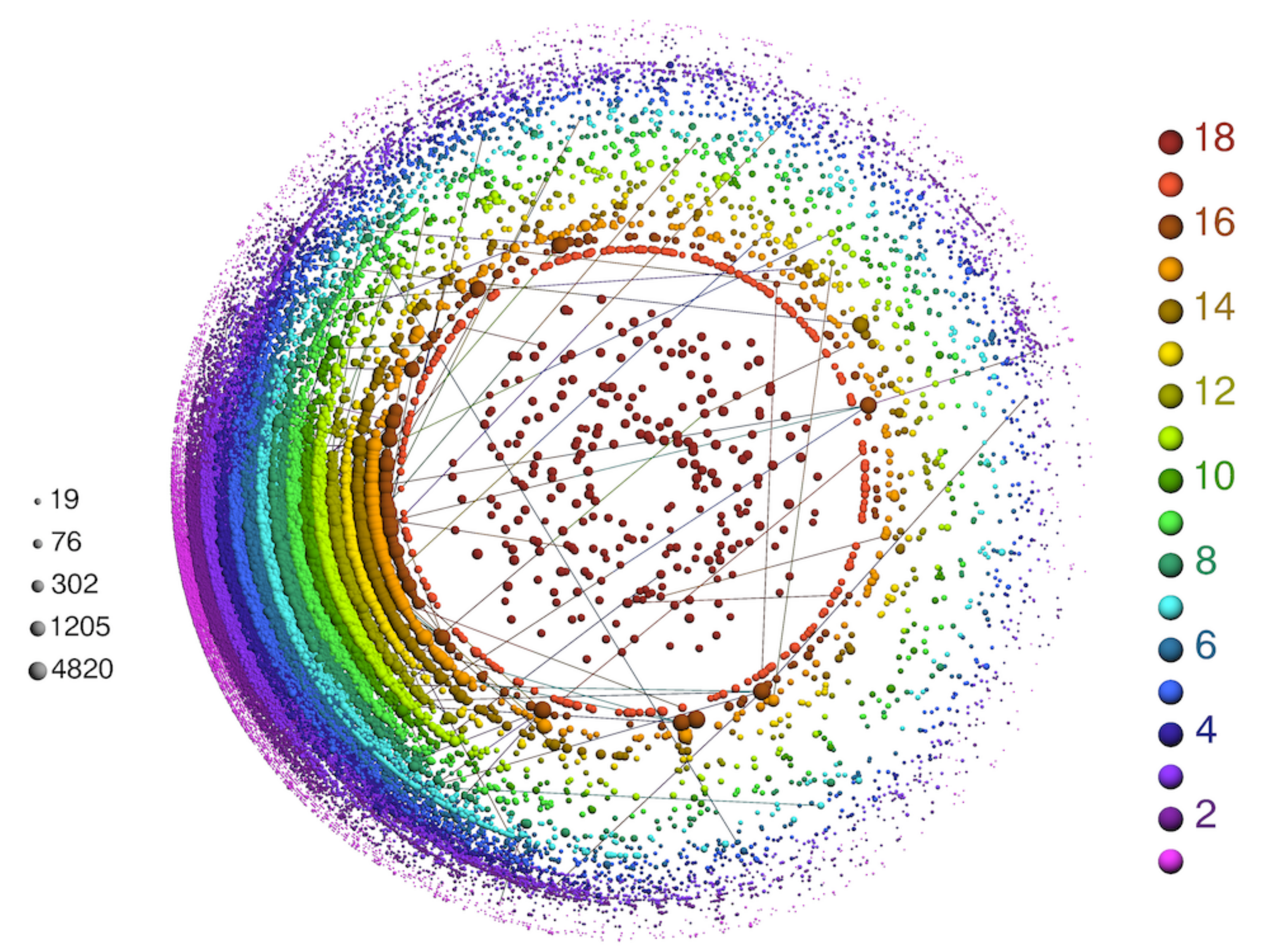}
\caption{Visualisation of the social network of active users, based on $k$-core decomposition and components analysis. The size of each vertex is proportional to its degree, whereas color codes the $k$-coreness. A sample of 10\% of the whole network has been used for this visualisation.}
\label{fig:corenet}
\end{figure}
In our analysis, we will distinguish between two different states for users in the social network: ``active'' and ``non-active'' vertices. 
We will indicate with ``tweeting activation'' or ``rumor spreading'' the user-to-user interaction process for transmitting information related to a particular topic. In the following, we will indicate with $A(t)$ and $D(t)$ the number of active and non-active users at time $t$, respectively, with $A(t)+D(t)=N$, where $N$ is total number of users considered in the social network. The observed social network of active users is shown in Fig.\,\ref{fig:corenet}, where a visualization based on $k$-core decomposition and component analysis is presented~\cite{alvarez2006large,beiro2008low}. The $k$-core of a graph is defined as the maximal connected subgraph in which all vertices have degree at least $k$. In practice, a $k$-core is obtained by recursively removing all vertices with degree smaller than $k$, until the degree of all remaining vertices is larger than or equal to $k$. The $k$-coreness of a vertex is the index of the highest $k$-core containing that vertex. Vertices with the highest $k$-coreness act as the most influential spreader of information in the network. In fact, it has been recently shown that in some plausible circumstances the best spreaders are not the most highly connected or the most central people but those with higher $k$-coreness \cite{kitsak2010identification}, and there is evidence of a positive correlation between $k$-coreness and the size of cascades of messages, suggesting that users at the core of the network are more likely to be the seeds of global chains of information diffusion \cite{gonzalez2011dynamics}.

The $k$-core decomposition allows to identify some salient features of the observed social network of active users, uncovering structural properties due to its specific topology. In Fig.\,\ref{fig:corenet}, the presence of an inhomogeneous distribution of vertices in the shells is a signature of non-trivial correlations. Moreover, the presence of vertices with high degree in any $k$-shell, i.e., a very low correlation between degree and shell-index, indicates that hubs are likely to be found also in external shells, a behavior typical of networks without an apparent global hierarchical structure like the World Wide Web \cite{alvarez2006large,beiro2008low}.

\begin{figure}[!t]
\includegraphics[width=9cm]{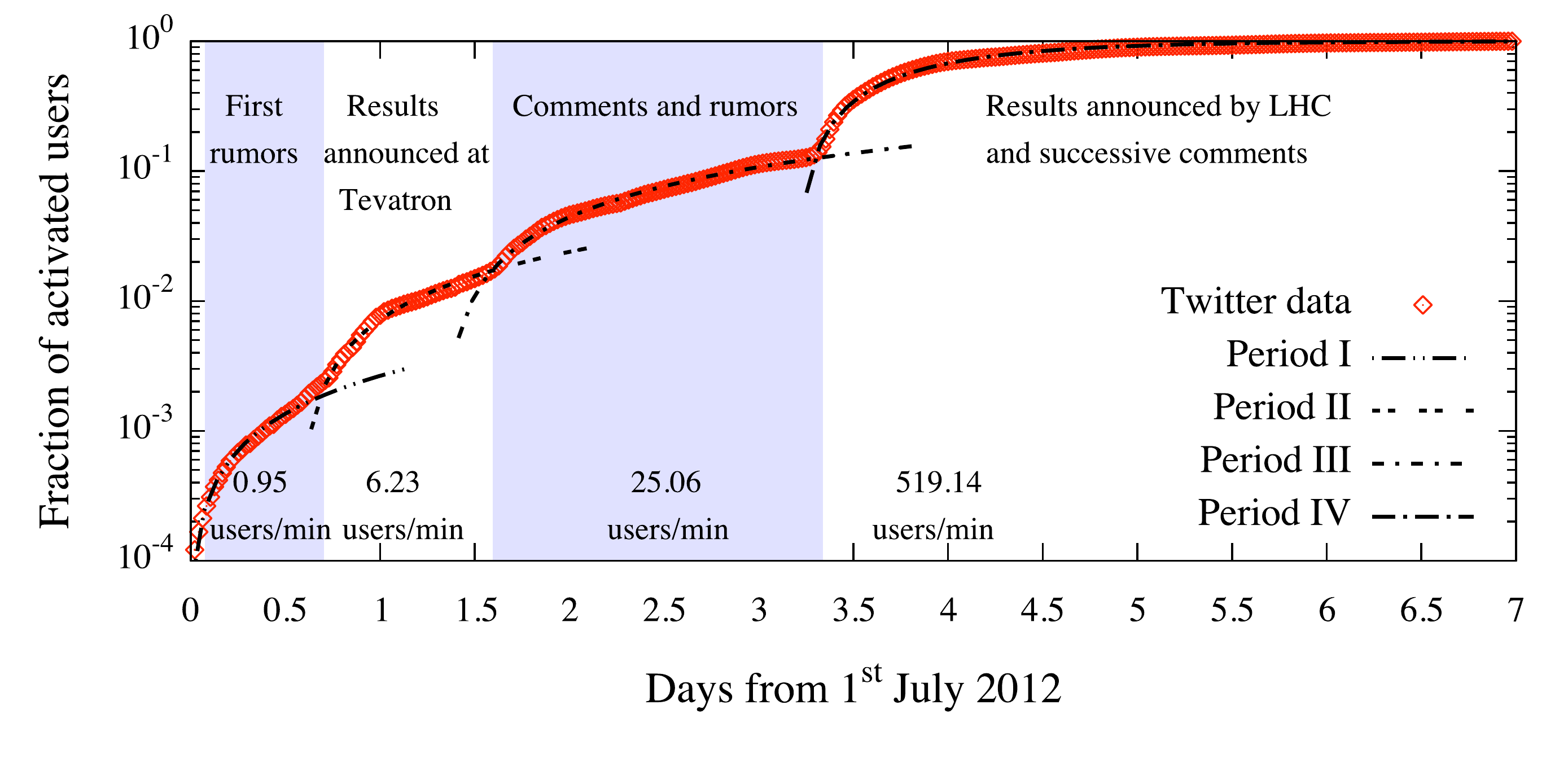}
\caption{Points indicate the fraction of users who are active at least once (see the text for more detail) with respect to the total number of users in the dataset at the end of the period taken into consideration, i.e., $A^{\star}(t =\text{8 July 2012})$, as a function of time. Lines indicate the fitting results obtained separately for each temporal range by adopting the model given by Eq.\,(\ref{def:simpleepidemicsol}). The rate of activation $\lambda^{\star}$ for each period is reported at the bottom of the figure.}
\label{fig:epidemics}
\end{figure}

\vspace{0.5truecm}\noindent\textbf{Modeling the Dynamics of User Activation without De-activation.} As the first step, we do not consider the influence of the structure of the network on the process.
We define a node as active at time $t$ if he or she has tweeted at least once about the Higgs boson within that instant of time. In the following, we indicate with $A^{\star}(t)$ and $a^{\star}(t)$ the number and the fraction of active users at time $t$, respectively.

Hence, the number $A^{\star}(t)$ of active users is expected to be a monotonic increasing function of time. We divide the whole period of data taking into four temporal ranges of interest, corresponding to periods I, II, III and IV, previously described. In Fig.\,\ref{fig:epidemics} we show for each period the observed evolution of the fraction $a^{\star}(t)=A^{\star}(t)/N$ of infected users versus time, where $N$ is the total number of users in the dataset in the data collection period.

In order to model the evolution of the active users over time, we firstly tried to exploit classic susceptible-infected (SI) models in an unstructured population~\cite{keeling2008modeling}, but this led to a very poor fit of the data.
For this reason, we developed a new model starting from the observation of the specific characteristics of our dataset.
In general, once a user has twitted, we observe that he or she will not probably tweet significantly about the Higgs boson in the near future, according to the bursty behavior shown previously. Therefore, we make the simplifying assumption that he or she will not tweet again after the tweeting activation. In this case, the number of newly active vertices at time $t$ is proportional to the number of users who have not been active before:

\begin{eqnarray}
\label{def:simpleepidemic} 
A^{\star}(t+\Delta t) = A^{\star}(t) + \lambda^{\star}\[N-A^{\star}(t)\] \Delta t,
\end{eqnarray}
where $\lambda^{\star}$ is a constant activation rate. In the limit of small $\Delta t$, we obtain the following ordinary differential equation:
\begin{eqnarray}
\label{def:simpleepidemicode} 
\frac{da^{\star}(t)}{dt}=\lambda^{\star}\[1-a^{\star}(t)\],
\end{eqnarray}
corresponding to our model for the fraction of users tweeting at least once about the Higgs boson. The evolution over time of $a^{\star}(t)$ is the solution of Eq.\,(\ref{def:simpleepidemicode}), given by
\begin{eqnarray}
\label{def:simpleepidemicsol} a^{\star}(t)=1-\[1-a^{\star}(t_{k})\]e^{-\lambda^{\star}(t-t_{k})},
\end{eqnarray}
where $k=$\,I, II, III and IV indicates the period of interest, $t_{k}$ is the starting date of period $k$ and $a^{\star}(t_{k})$ is the corresponding initial fraction of users active at least once.

We fit the evolution function given by Eq.\,(\ref{def:simpleepidemicsol}) to the observed data for each period of interest: the resulting model for each case is shown in Fig.\,(\ref{fig:epidemics}), demonstrating the agreement with the data. The activation rate increases during the four intervals of time taken into consideration, from about one user per minute on 1\superscript{st} July 2012, up to about 519 users per minute in the last period.

\vspace{0.5truecm}\noindent\textbf{Modeling the Dynamics of User Activation with De-activation.}\label{sec-useract} In this subsection, we will focus on the propagation of interest on the event through social cascading. A user is considered non-active in a given time window $\Delta t$ if he or she has not tweeted in that time interval. In other words, in this refined model, an active user can become non-active again (de-activated) if he or she does not keep tweeting about the Higgs boson. 
In the time interval between $t$ and $t+\Delta t$ active users can become non-active after a certain amount of time for any reason: we indicate with $\beta(t)$ the probability per unit of time for the transition from active to non-active state. Hence, the number of users that becomes non-active in the interval $\Delta t$ is given by $\beta(t)A(t)\Delta t$.
By introducing de-activation we also account for the limited visibility of tweets on the timelines of Twitter clients, i.e., newer tweets replace older ones. Moreover, we observe that the number of non-active users at time $t$ that will become active at time $t+\Delta t$ is a function of both their in-going degree and the out-going degree of active users  at time $t$. 

A non-active user connected to more than one active user at the same time is more likely to become active with respect to non-active users connected to only one active user. Let us indicate with $j_{A}$ the number of active users connected to a non-active user. If $\lambda(t)$ indicates the activation probability per unit of time per link, for a non-active user with degree $k^{in}$ the probability per unit of time of changing from non-active to active state is given by $p_{\lambda}(t;j_{A})=1-\[1-\lambda(t)\]^{j_{A}}$. In general, the probability that such a non-active user is connected to $j_{A}$ active users at the same time depends on the out-going degree of active users, i.e., on network vertex-vertex correlations. More specifically, such a probability depends on the conditional probability of observing a vertex with out-going degree $k^{out}$ connected to a vertex with in-going degree $k^{in}$.

It has been shown that a pure scale-free degree distribution with exponent between 2 and 3 is a sufficient condition for the absence of an epidemic threshold in unstructured networks with arbitrary two-point degree correlation function \cite{boguna2003absence}, i.e., correlations at neighborhood level do not affect the spreading dynamics.
We use this result as a simplifying assumption for modeling the spreading in our network, exhibiting a scale-free degree distribution with exponent 2.5 for $k>200$. Therefore, we neglect correlations and we estimate the probability that a non-active user, with in-going degree $k^{in}$, is connected to $j_{A}$ active users, with \emph{any} out-going degree, by
\begin{eqnarray}
\label{def:probconn}
\tilde{p}(t;j_{A},k^{in}) = \frac{\binom{A(t)}{j_{A}}\binom{N-A(t)-1}{k^{in}-j_{A}}}{\binom{N-1}{k^{in}}},
\end{eqnarray}
accounting for all the possible ways to arrange $A(t)$ activations within $j_{A}$ users from the total number of possible combinations of the remaining $N-1$ users within $k^{in}$ users.

Hence, the probability that a non-active user with in-going degree $k^{in}$ is activated by at least one active user in its neighborhood is given by
\begin{eqnarray}
P_{\lambda,k^{in}}(D\lto A)=\sum_{j_{A}=1}^{k^{in}}\tilde{p}(t;j_{A},k^{in})p_{\lambda}(t;j_{A}).
\end{eqnarray}
It follows that the total probability that non-active users will become active per unit of time is given by
\begin{eqnarray}
\label{def:probact}\Theta_{\lambda}(t) = \sum_{k^{in}}\mathcal{P}(k^{in}) P_{\lambda,k^{in}}(D\lto A),
\end{eqnarray}
being $\mathcal{P}(k^{in})$ the probability density of the in-going degree. It follows that $(N-A(t))\mathcal{P}(k^{in})$ indicates the number of non-active users with in-going degree $k^{in}$ at time $t$. We model the dynamics of the number of active users in the time interval $\Delta t$ by
\begin{eqnarray}
\label{def:epidemic} A(t+\Delta t) = A(t) + \[ -\beta(t)A(t) + (N-A(t))\Theta_{\lambda(t)}(t) \]\Delta t.\nonumber
\end{eqnarray}
Therefore, by choosing $\Delta t=1$, i.e., equal to the time unit of observation, we obtain the general discrete model
\begin{eqnarray}
\label{def:epidemicode} A(t+1) = \(1 - \tilde{\beta}(t)\)A(t) + (N-A(t))\Theta_{\tilde{\lambda}(t)}(t),
\end{eqnarray}
valid for the general case of activation and de-activation rates that change over time. 

In Eq.\,(\ref{def:epidemicode}) the parameters $\tilde{\beta}=\beta\Delta t$ and $\tilde{\lambda}=\lambda\Delta t$ indicate probability instead of probability rates.
However, in the particular case of $\Delta t=1$ it is possible to mix rates and probabilities because both will have the same values, even though their units are different \cite{gomez2011nonperturbative}: for sake of simplicity, in the following we use the notation $\beta=\tilde{\beta}$ and $\lambda=\tilde{\lambda}$. Eq.\,(\ref{def:epidemicode}) represents the balance equation indicating that the number of active users at a certain instant is given by the number of vertices that at the previous instant did not change from active to non-active state plus the number of newly active users. In the following we will consider the density of active users defined by $\rho(t)=A(t)/N$, leading to the evolution equation
\begin{eqnarray}
\label{def:epidemicrho} \rho(t+1) = \(1 -\beta(t)\)\rho(t) + (1-\rho(t))\Theta_{\lambda(t)}(t).
\end{eqnarray}

In general, the solution of Eq.\,(\ref{def:epidemicode}) and Eq.\,(\ref{def:epidemicrho}) cannot be obtained analytically because of the complexity of $\Theta_{\lambda(t)}(t)$: therefore, some simplifying assumptions or numerical methods should be adopted instead.

\begin{figure*}[!t]
\includegraphics[width=\textwidth]{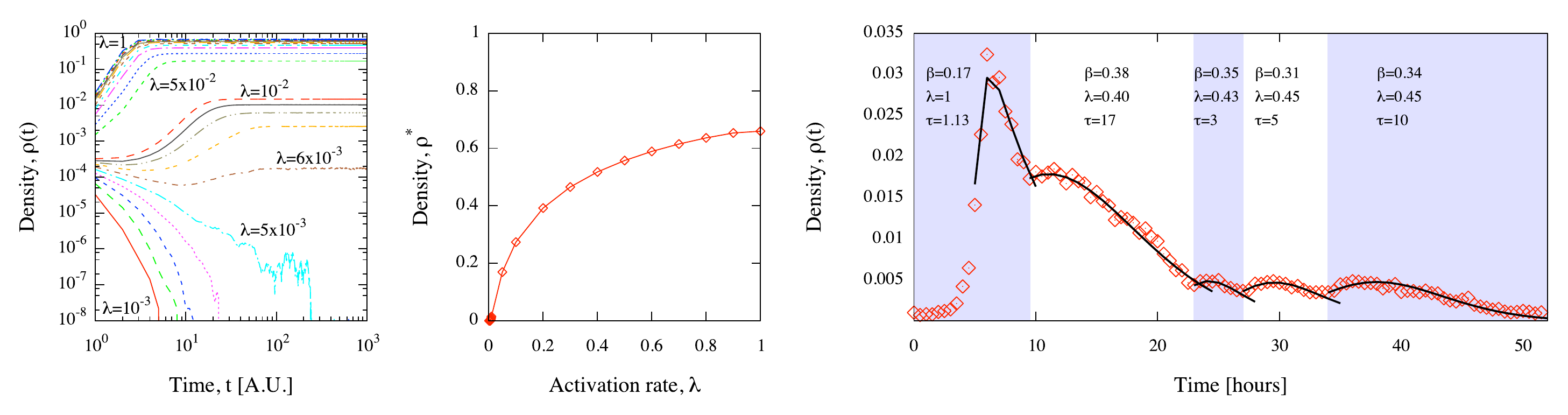}
 \caption{\textbf{First panel:} Evolution of the density of active users versus time obtained from simulations of spreading dynamics. The de-activation rate is $\beta=1$ and the different curves correspond to different values of the activation rate $\lambda$. Each curve corresponds to the ensemble average of 200 random independent realizations. \textbf{Second panel:} Average value of the density of active users in the stationary state, as a function of $\lambda$. \textbf{Third panel:} Observed evolution of the density of active users versus time (points) in Period IV, i.e., during and after the main event, from 03:00 AM, 4\superscript{th} July. Curves indicate the predictions obtained from the model defined by Eq.\,\ref{def:epidemicrho} coupled to Eq.\,\ref{def:epidemicrho2}, where the values of the corresponding parameters are reported in the figure for different sub-periods. The reported $\lambda$ refers to the initial value of the activation rate.}
 \label{fig:epidemics-evo}
\end{figure*} 

Let us focus only on Period IV, i.e., during and after the main event, from 03:00 AM on 4\superscript{th} July to the end of the data collection period. The initial fraction of active users is approximately $\rho(0)=0.1\%$ of the total number of users in our dataset.

In order to assess the validity of our analytical model, we perform large-scale Monte Carlo simulations of the spreading dynamics through the network of observed connections among users. More specifically, we consider the case where activation and de-activation rates do not change over time: we vary their values from 0 to 1, independently; for each possible configuration corresponding to the pair $(\beta,\lambda)$ we perform 200 random independent realizations of rumor spreading and we calculate the ensemble average at each time step $t$ to obtain an estimation of the expected value of the density $\rho(t)$. The results for the case with $\beta=1$ and several different values of $\lambda$ are shown in the first two panels of Fig.\,\ref{fig:epidemics-evo}. In the first panel, we show the evolution of $\rho(t)$ versus time: for $\lambda<6\times10^{-3}$ the density $\rho(t)$ tends to decrease to zero for increasing time, while for $\lambda\geq 6\times10^{-3}$ the density $\rho(t)$ tends to reach a stationary state, indicating that the spreading becomes endemic. We indicate with $\rho^{\star}$ the stationary value reached by $\rho(t)$ after the transient time. In the second panel, we show the value of $\rho^{\star}$ versus the activation rate: the endemic state, where $\rho^{\star}>0$, is quickly reached for small values of $\lambda$. This result is qualitatively confirmed by our analytical model (see Eq.\,(\ref{def:epidemicrho})) and it is in agreement with the result reported in~\cite{boguna2003absence}, stating that the epidemic threshold of an endemic state tends to zero for increasing network size with a scale-free topology. However, such results do not reproduce the observed spreading dynamics, whose density $\rho(t)$ is shown in the third panel of Fig.\,\ref{fig:epidemics-evo}. The data show a quickly increasing number of users within a few hours, with a maximum value reached at the beginning of the International Conference on High-Energy Physics. Such a fast increasing behavior can be explained by tweets related to the excitement for a possible announcement of the discovery of the Higgs boson. In fact, the number of active users  in the following hour rapidly decreases by about 40\%, staying stable for the subsequent 2 hours and then decreasing again. 

In case of epidemics with constant activation rate in scale-free networks with a large number of nodes we expect the appearance of an endemic state. However, this is not the case in our dataset. For this reason, we modify the model by introducing a variable activation rate $\lambda$, accounting for the decreasing interest on a tweet over time, according to recent studies suggesting the existence of a natural time scale over which attention fades \cite{wu2007novelty}.
We model the evolution of $\lambda$ as follows:
\begin{eqnarray}
\label{def:epidemicrho2} 
\lambda(t+1)&=& \(1-\xi\)\lambda(t),\qquad 0<\xi<1,
\end{eqnarray}
which is the discrete counterpart of the continuous equation whose solution is the exponential decay $\lambda(t)=\lambda(t_{0}) e^{-\xi (t-t_{0})}$. Here, we interpret $\xi$ as the inverse of a characteristic scale $\tau$ regulating the decay dynamics. We use the coupled equations (\ref{def:epidemicrho}) and (\ref{def:epidemicrho2}) to model the observed spreading dynamics. 

During the whole Period IV, we identify five sub-periods, each one characterized by an increasing number of active users followed by a decreasing one. We then vary the parameters $\beta$, $\lambda$ and $\tau$ in order to try to reproduce the data in each sub-period. The solid curves in the third panel of Fig.\,\ref{fig:epidemics-evo} correspond to our model (Eq.~(\ref{def:epidemicrho}) and Eq.~(\ref{def:epidemicrho2})) using the set of parameters minimizing $\chi^{2}$. The rapid increase of active users in the first sub-period of Period IV is followed by a fast decrease, with time scale $\tau\approx 1.13$\,hours, initial activation rate $\lambda_{0}=1$ and $\beta=0.17$. Such a fast decreasing trend is slowed after about 9 hours, approximately at the time when the rumor has reached the other side of the world in the early morning: from this time instant up to the end of the observation, the values of de-activation probability and the initial value of the activation probability are almost constant (ranging from 0.31 to 0.38, and from 0.40 to 0.45, respectively). In the following sub-periods only the decay time scale $\tau$ significantly varies from 3 to 17 hours.

\section*{Discussion}

On 4\superscript{th} July 2012, the ATLAS and CMS collaborations announced the discovery of a new particle, with the same features of the elusive Higgs boson. Such a finding represents a milestone in particle physics and a unique occasion to study the dynamics of information spreading on a global scale.
In this study, we have monitored user activities on Twitter before, during and after the announcement of the discovery of this Higgs boson-like particle.

The joint analysis of spatial and temporal user activity patterns unveiled specific dynamics in different periods of the rumor spreading. Before the announcement of the teams based at CERN, tweets were more likely to be sent within a few seconds by users living within 20\,km. During the main event the activity became frenetic and its time scale reduced to 2 seconds without a specific spatial pattern. After the main event a less frenetic activity has been observed while users from any part of the world were still involved in the process, with no apparent prevalence of small or large inter-tweets spaces.

Finally, we have focused our attention on the network of individuals who posted at least one message about the discovery (tweets, re-tweets or reply to a tweet). The observed network of users exhibits a non-trivial structure with a typically bursty tweeting process happening over it.
We have proposed a model for information spreading with variable activation rate in heterogenous networks showing that we are able to reproduce the collective behavior of about 500,000 users with remarkable accuracy. Our model assumes memoryless individuals where the activation process is driven by social reinforcement at neighborhood level. The active (or non-active) states of his or her social neighbors at each time step act as a ``topological memory'', causing the individual to be activated with larger probability if most of his or her friends are tweeting the rumor repeatedly in time.

Even if the proposed models have been developed for this specific series of events, we believe that the proposed framework can be applied and fitted to many other spreading processes on online and physical social networks.

\section*{Methods}

For each search, Twitter provides information about the number of missing tweets, which is usually negligible. In any case, it worth noting that during this data collection no messages related to missing tweets were received: for this reason, to the best of our knowledge, we might claim that this dataset includes \textit{all} the tweets satisfying our search criteria.

Our original list of relevant keywords was larger, including terms as \texttt{alice} and \texttt{cms}.
However, the amount of tweets retrieved using these keywords but not related to the Higgs boson was not negligible and, for this reason, we decided to avoid considering such terms in our analysis.

The geographic names contained in the \textit{location} textfield of Twitter user profiles were converted using the Google Geocoder API.

Recent studies make use of the retweet network (i.e., who retweets whom) and the mentions network (i.e., who mentions whom) to investigate, for instance, the patterns of sentiment expression~\cite{bliss2012twitter}. However, it is worth remarking that the main source of information consumption on Twitter website and clients is the home timeline, which contains messages from all social contacts, regardless of the number of reciprocal interactions. Therefore, users activity can be triggered (or not) according to their interests. In our study, we prefer to use the follower network to investigate the temporal dynamics of the fraction of people involved in the process of spreading the information about the Higgs boson discovery as a function of time.

The social graph was obtained by retrieving the following list for each user participating in the process. We make the reasonable assumption that the time required to collect the data is much shorter than the time required to observe significant changes in the full social graph.

\begin{acknowledgments}
The authors thank J.I. Alvarez-Hamelin and A. Mambrini for useful and fruitful discussions. This work was supported through the EPSRC Grant ``The Uncertainty of Identity: Linking Spatiotemporal Information Between Virtual and Real Worlds'' (EP/J005266/1).
\end{acknowledgments}

\addcontentsline{toc}{section}{References} 

\bibliographystyle{naturemag}
\bibliography{draft}

\end{document}